\documentclass[conference]{IEEEtran}
\IEEEoverridecommandlockouts

\usepackage{cite}
\usepackage{amsmath,amssymb,amsfonts}
\usepackage{algorithmic}
\usepackage{graphicx}
\usepackage{textcomp}
\usepackage{xcolor}
\usepackage{comment}
\usepackage[acronym]{glossaries}
\usepackage[free-standing-units,per-mode=repeated-symbol,binary-units=true,detect-weight=true,detect-family=true]{siunitx}[=v2]
\usepackage{lscape}
\usepackage{multirow}

\usepackage{booktabs}
\usepackage{tabularx}
\usepackage{array}
\usepackage{makecell}
\usepackage[table]{xcolor}
\usepackage{url}

\usepackage{pifont}

\usepackage{tikz}
\definecolor{ethblue}{HTML}{215CAF}
\definecolor{ethred}{HTML}{B7352D}
\definecolor{ethgreen}{HTML}{627313}
\definecolor{ethpetrol}{HTML}{007894}
\definecolor{ethbronze}{HTML}{8E6713}
\definecolor{ethpurple}{HTML}{A7117A}

\newcommand*\circledletter[2][blue]{%
  \tikz[baseline=(char.base)]{
    \node[
      shape=circle,
      fill=#1,
      text=white,
      inner sep=0.8pt
    ] (char) {#2};}}

\usepackage{threeparttable} 
\usepackage{booktabs}  

\newcommand*\circled[1]{\tikz[baseline=(char.base)]{
            \node[shape=circle,draw,inner sep=1pt] (char) {#1};}}

\def\BibTeX{{\rm B\kern-.05em{\sc i\kern-.025em b}\kern-.08em
    T\kern-.1667em\lower.7ex\hbox{E}\kern-.125emX}}

\usepackage[most]{tcolorbox}

\newtcolorbox{insightbox}{
  colback=gray!12,
  colframe=gray!12,
  boxrule=0pt,
  arc=2pt,
  left=1pt,
  right=1pt,
  top=1pt,
  bottom=1pt
}

\usepackage{threeparttable}

\begin{document}

\title{At-the-Roofline Sparse Tensor Contractions on Vector Processors for Transformer Inference \\
}

\author{
\IEEEauthorblockN{
Bowen Wang\textsuperscript{*},
Chi Zhang\textsuperscript{*},
Diyou Shen\textsuperscript{*},
Renzo Andri\textsuperscript{\textdaggerdbl},
Navaneeth Kunhi Purayil\textsuperscript{*},
Luca Benini\textsuperscript{*\textdagger}
}

\IEEEauthorblockA{
\textsuperscript{*}IIS, ETH Zürich, Switzerland;
\textsuperscript{\textdagger}DEI, Università di Bologna, Italy;
\textsuperscript{\textdaggerdbl}Computing Systems Lab, Huawei, Switzerland
}

\IEEEauthorblockA{
\{bowwang, chizhang, dishen, nkunhi, lbenini\}@iis.ee.ethz.ch,
renzo.andri@huawei.com
}
\vspace{-0.5em}
\thanks{This work is supported by the ETH Future Computing Laboratory (EFCL) and Huawei ZRC.}
\vspace{-2em}
}

\newacronym{LLM}{LLM}{Large Language Model}
\newacronym{ML}{ML}{Machine Learning}
\newacronym{MS}{MS}{moderate-sparsity}

\newacronym{NPU}{NPU}{Neural Processing Unit}
\newacronym{FPU}{FPU}{Floating-Point Unit}
\newacronym{SIMD}{SIMD}{Single Instruction Multiple Data}
\newacronym{VLEN}{VLEN}{vector length}
\newacronym{VPE}{VPE}{Vector Processing Element}
\newacronym{VAU}{VAU}{Vector Arithmetic Unit}
\newacronym{VLSU}{VLSU}{Vector Load-Store Unit}
\newacronym{VRF}{VRF}{Vector Register File}
\newacronym{RVV}{RVV}{RISC-V Vector}
\newacronym{CC}{CC}{Core Complex}
\newacronym{TCDM}{TCDM}{Tightly Coupled Data Memory}
\newacronym{CSR}{CSR}{Control and Status Register}
\newacronym{SPM}{SPM}{Scratchpad Memory}
\newacronym{SCM}{SCM}{standard-cell memory}

\maketitle

\begin{abstract}

Fine-grained weight pruning and activation sparsification have emerged as effective approaches for reducing the compute and memory cost of inference for Transformer models. In the moderate-sparsity regime, Gustavson's dataflow provides a natural execution model for exploiting both activation and weight sparsity on vector processors through metadata-driven indexed accumulation. However, existing \gls{RVV} architectures lack native support for this pattern, forcing kernels to rely on software index decoding and L1-backed indexed memory operations that keep sparse tensor contractions far below their roofline performance bound. We present Ventaglio, a runtime-configurable sparse execution unit coupled with \gls{RVV} ISA extensions that drives sparse tensor contractions toward their roofline through indexed gather-accumulate-scatter support. Integrated into an open-source vector processing cluster and implemented in 12\,nm FinFET, Ventaglio accelerates sparse tensor contraction kernels by $6.9\text{--}7.4\times$ over optimized \gls{RVV} baselines, with only $3.1\%$ area overhead for a cluster of tightly-L1 coupled vector processing elements. We build a performance-accurate instruction-level model of the Ventaglio extension, calibrate it against RTL implementation, and leverage it for scale-out performance analysis on a large $4\times4$ multi-cluster system. Using a DuoGPT-pruned LLaMA-3-8B model with practical $40\text{--}60\percent$ dual sparsity, Ventaglio achieves $2.40\text{--}5.25\times$ and $2.06\text{--}3.16\times$ speedup over dense baselines during prefill and autoregressive decoding, respectively.

\end{abstract}

\begin{IEEEkeywords}
Vector processor, RISC-V, sparse tensor contraction, roofline analysis
\vspace{-1em}
\end{IEEEkeywords}

\glsresetall
\bstctlcite{IEEEexample:BSTcontrol}

\section{Introduction}

The success of Transformer architectures is closely tied to scaling laws, motivating the development of models with billions to trillions of parameters\cite{scaling_law}. However, the resulting growth in model size incurs substantial memory footprint and computational cost, limiting efficient deployment on edge platforms with tight memory and bandwidth budgets. 

Sparsity has been exploited to mitigate these costs. Fine-grained model pruning statically induces \textit{weight sparsity} by removing redundant parameters~\cite{yang2025wanda++, liu2025toward, bai2024sparsellm}, whereas \textit{activation sparsity} arises dynamically during inference and is exploited via gating and zero-skipping~\cite{liu2024training, akhauri2024shadowllm, ma2024first}.
In this work, we focus on the \gls{MS} regime, corresponding to a range of zero-element ratios of 40--75\%, which commonly arises in practical Transformer inference. Translating these sparsity into practical speedups requires tight hardware--software co-design.

Existing hardware support remains largely matrix-centric. NVIDIA extends Tensor Core MMA pipelines with 2:4 sparse operand support~\cite{nvidia2020ampere, nvidia2022hopper, nvidia2025blackwelltc}, AMD provides sparse-matrix ISA support on Instinct accelerators~\cite{amd2025mi300isa}, and Arm has introduced analogous support in SME2\cite{arm2025sme2_sparse_op}, while \glspl{NPU} such as Ethos-U rely on dedicated weight-decoding hardware~\cite{arm2024ethosu_arch}. These approaches map well to the \textit{prefill} stage of Transformer inference, where computation retains high matrix-level parallelism. During \textit{autoregressive decoding}, however, execution shifts toward lower-intensity matrix-vector operations, under which such engines are more easily underutilized.

\begin{figure*}[ht]
    \vspace{-2.2em}
  \centering
  \includegraphics[width=\linewidth]{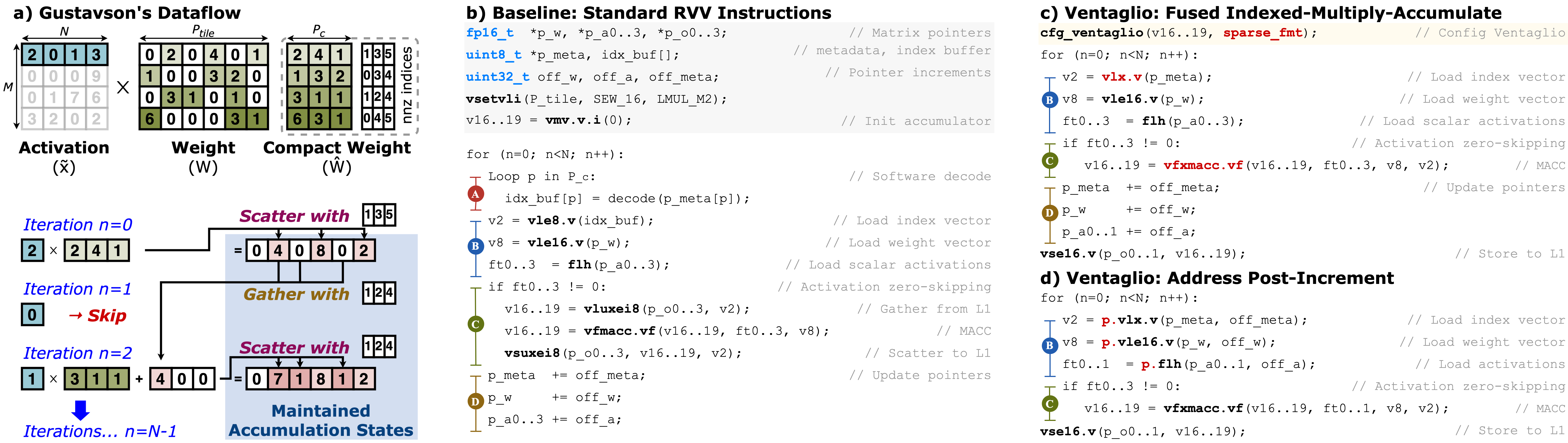}
  \vspace{-2.3em}
  \caption{(a) Gustavson's algorithm for sparse tensor contraction. (b) Baseline realization with standard RVV instructions. (c) Ventaglio kernel implementation with fused indexed-multiply-accumulate extensions. (d) Ventaglio address post-increment extension for streamlined sparse updates.}
  \label{fig:kernels}
    \vspace{-1em}
\end{figure*}

Vector processors are a compelling substrate for such workloads, offering a favorable balance of performance, energy efficiency, and programmability\cite{dabbelt2016vector}. In particular, Gustavson's dataflow provides a natural execution model for combining fine-grained weight sparsity with dynamic activation sparsity: as depicted in Fig. \ref{fig:kernels}(a), non-zero activations trigger computation, while weight metadata determines the indexed output accumulations to perform. These properties make vector processors attractive for scalable \gls{ML} systems, where lightweight \glspl{VPE} are organized into tightly coupled clusters with high-throughput interconnects to shared L1 memory\cite{perotti2025spatz}, enabling further hierarchical scale-out across multiple clusters.

However, existing \gls{RVV} architectures provide no native support for this metadata-driven gather/accumulate/scatter pattern, leaving optimized kernel libraries to rely on explicit software metadata decoding and L1-backed indexed memory operations. The recently proposed IndexMAC~\cite{titopoulos2025optimizing} and SCG~\cite{tengfei2026scg} \gls{RVV} ISA extensions partially reduce this overhead through indexed access to \gls{VRF}-resident matrix tiles, but remain tied to matrix-level tiling and reuse, leaving vector-matrix execution and activation-driven sparsity insufficiently addressed.


\definecolor{lightred}{RGB}{252,228,228}
\definecolor{lightgreen}{RGB}{226,239,218}
\definecolor{lightyellow}{RGB}{255,242,204}

\newcommand{\cmark}{\ding{51}}
\newcommand{\xmark}{\ding{55}}

\newcommand{\yescell}{\cellcolor{lightgreen}\cmark}
\newcommand{\nocell}{\cellcolor{lightred}\xmark}
\newcommand{\swcell}{\cellcolor{lightyellow}SW}
\newcommand{\hwswcell}{\cellcolor{lightgreen}\makecell[c]{HW,\\SW}}
\newcommand{\dashcell}{\cellcolor{lightred}--}
\newcommand{\partcell}{\cellcolor{lightyellow}\textbf{$\sim$}}

\renewcommand\theadfont{\bfseries}
\renewcommand\theadalign{cc}
\renewcommand\theadgape{}

\begin{table}[t]
\centering
\scriptsize
\setlength{\tabcolsep}{2.5pt}
\renewcommand{\arraystretch}{1.15}

\begin{threeparttable}
\caption{Comparison with fine-grained sparse acceleration designs.}
\vspace{-1em}
\label{tab:soa_comparison}

\begin{tabular}{
    >{\raggedright\arraybackslash}m{2.20cm}
    >{\centering\arraybackslash}m{0.75cm}
    >{\centering\arraybackslash}m{0.75cm}
    >{\centering\arraybackslash}m{0.75cm}
    >{\centering\arraybackslash}m{0.75cm}
    >{\centering\arraybackslash}m{0.75cm}
    >{\centering\arraybackslash}m{0.75cm}
    >{\centering\arraybackslash}m{0.75cm}
}
\toprule
\textbf{\rule{0pt}{2.4ex}Design} &
\textbf{Arch.} &
\textbf{Dataflow} &
\textbf{Fmt.} &
\makecell[c]{\textbf{Mat-Vec}\\\textbf{Eff.}} &
\makecell[c]{\textbf{Mat-Mat}\\\textbf{Eff.}} &
\makecell[c]{\textbf{Act.}\\\textbf{Sparsity}} &
\makecell[c]{\textbf{Open}\\\textbf{Source}} \\
\specialrule{\lightrulewidth}{0.5pt}{0.5pt}

\textbf{NVIDIA Blackwell}\cite{nvidia2025blackwelltc} & GPU & IP    & 2:4 & \nocell  & \yescell & \nocell & \swcell \\
\specialrule{\lightrulewidth}{0.5pt}{0.5pt}
\textbf{AMD MI300}\cite{amd2025mi300isa}              & GPU & IP    & 2:4 & \nocell  & \yescell & \nocell & \swcell \\
\specialrule{\lightrulewidth}{0.5pt}{0.5pt}
\textbf{Arm C1-Ultra (SME2)}\cite{arm2025sme2_sparse_op} & CPU & OP & 2:4 & \partcell & \yescell & \nocell & \swcell \\
\specialrule{\lightrulewidth}{0.5pt}{0.5pt}
\textbf{Arm Ethos-U85}\cite{arm2024ethosu_arch}       & NPU & IP    & 2:4 & \nocell  & \yescell & \nocell & \dashcell \\
\specialrule{\lightrulewidth}{0.5pt}{0.5pt}

\textbf{IndexMAC}\cite{titopoulos2025optimizing} & \gls{VPE} & Gust. & N:M & \nocell  & \yescell & \nocell & \dashcell \\
\specialrule{\lightrulewidth}{0.5pt}{0.5pt}
\textbf{SCG}\cite{tengfei2026scg} & \gls{VPE} & OP & Custom. & \partcell  & \yescell & \nocell & \dashcell \\
\specialrule{\lightrulewidth}{0.5pt}{0.5pt}
\makecell[l]{\textbf{Ventaglio}\\(This Work)} & \gls{VPE} & Gust. & \makecell[c]{N:M,\\Bitmap} & \yescell & \yescell & \yescell & \hwswcell \\
\bottomrule
\end{tabular}

\vspace{-0.4em}

\end{threeparttable}
\vspace{-2.0em}
\end{table}

To address these limitations, we present \textit{Ventaglio}, a runtime-configurable sparse execution unit for vector processors that elevates metadata-driven indexed accumulation to a native \gls{RVV} execution primitive. Ventaglio couples the \gls{VAU} with a multi-channel memory unit that maintains expanded data structures outside the \gls{VRF} and materializes only the referenced lanes required by each sparse update, thereby sustaining full \gls{VAU} throughput. 
Its modular scatter/gather datapath further enables a unified execution framework across different sparse encoding schemes. 
The contributions of this paper are:
\begin{itemize}
    \item We present the Ventaglio microarchitecture and the coupled \gls{RVV} ISA extensions tailored to metadata-driven sparse execution, providing index loading, indexed multiply-accumulate, and address-post-increment semantics for scalar and vector memory operations to reduce control overhead in sparse kernels\footnotemark.
    \item We integrate Ventaglio into the open-source dual-core Spatz cluster\cite{perotti2025spatz} and demonstrate $6.9\text{--}7.4\times$ speedup over the optimized RVV kernel baselines at $89.3-97.4\%$ \gls{VAU} utilization on \gls{MS} tensor contraction kernels. Implemented in 12\,nm FinFET technology, the extension incurs only {$3.1\%$} cluster-level area overhead.
    \item We build a performance-accurate instruction-level model of the Ventaglio extension in the GVSoC simulation framework \cite{bruschi2021gvsoc}, calibrate it against RTL implementation, and use it to evaluate end-to-end scale-out inference on a $4\times4$ multi-cluster system running DuoGPT--pruned~\cite{yin2025duogpt} LLaMA-3-8B~\cite{grattafiori2024llama} model. With 40-60\% dual--sparsity, Ventaglio achieves 2.06\text{--}5.25$\times$ speedup over dense inference across prefill and autoregressive decoding.
\end{itemize}

\footnotetext{Available at: \textcolor{blue}{\url{https://github.com/pulp-platform/spatz}}}
\section{Background and Motivation}

\subsection{Gustavson's Dataflow for Sparse Tensor Contraction}

We consider the multiplication of a gated activation vector with a weight matrix stored in a compact sparse format. Let $\tilde{\mathbf{x}} = g(\mathbf{x}) \in \mathbb{R}^{N}$
denote the activation vector after gating, $\mathbf{W} \in \mathbb{R}^{N \times P_{tile}}$ denote the logical weight matrix, and $\hat{\mathbf{W}} \in \mathbb{R}^{N \times P_c}$ denote its compact representation.
The target computation is defined over the logical matrix as
\begin{equation}
\vspace{-0.6em}
    \mathbf{y} = \tilde{\mathbf{x}} \mathbf{W},
\vspace{-0.3em}
\end{equation}
where $\mathbf{y} \in \mathbb{R}^{P_{tile}}$ is the dense output vector, while $\hat{\mathbf{W}}$ together with its metadata defines the stored representation used by the hardware. This formulation generalizes naturally to matrix--matrix contraction by processing multiple activation rows in parallel, i.e., loop unrolling along the M dimension.

As depicted in Fig.~\ref{fig:kernels}(a), for each nonzero activation $x_i$, the corresponding compressed row $\mathbf{W}_{i,:}$ is fetched and used to update a sparse subset of the output accumulator. Denoting by $\mathcal{I}_i$ the set of output coordinates referenced by the compact metadata of row $i$, the update can be written as
\begin{equation}
\vspace{-0.6em}
    y_j \leftarrow y_j + x_i \cdot w_{i,j}, \quad \forall j \in \mathcal{I}_i.
\vspace{-0.3em}
\end{equation}
Hence, activation sparsity determines \emph{when} useful computation is triggered, while weight metadata determines \emph{where} the corresponding partial sums must be accumulated.

Fig.~\ref{fig:kernels}(b) illustrates the baseline realization of this dataflow using standard RVV instructions. Concretely, each sparse update consists of:
\circledletter[ethred]{A} decoding the metadata into explicit indices compatible with RVV indexed operations;
\circledletter[ethblue]{B} loading the activation, weight values, and indices;
\circledletter[ethgreen]{C} executing gather-accumulate-scatter through indexed load/store instructions; and
\circledletter[ethbronze]{D} explicit pointer updates.
While functionally correct, this realization exposes three fundamental inefficiencies.

Standard RVV provides indexed memory operations for irregular accesses, but it does not natively interpret compressed sparse metadata. Therefore, software must first decode the sparse representation into explicit output indices before RVV can perform the corresponding gather and scatter operations, placing metadata interpretation on the software critical path.

\begin{insightbox}
\vspace{-0.4em}
\textit{Insight} \circledletter[black]{1}. \textit{Sparse metadata interpretation should be supported as a native hardware mechanism rather than reconstructed repeatedly in software.}
\vspace{-0.4em}
\end{insightbox}

Additionally, indexed accumulation is performed through generic indexed memory operations against the L1 memory system. Therefore, gather and scatter imply memory traffic rather than being handled more efficiently as core-local sparse update semantics. This inflates access overhead and increases pressure on the memory hierarchy.
\begin{insightbox}
\vspace{-0.4em}
\textit{Insight} \circledletter[black]{2}. \textit{Indexed gather/scatter for sparse accumulation should be handled in-core.}
\vspace{-0.4em}
\end{insightbox}

Moreover, the baseline flow materializes both the gathered accumulator elements and the updated sparse results as explicit vectors in \gls{VRF}. This loosens the coupling between data movement and computation, limiting efficient vector chaining across sparse updates.
\begin{insightbox}
\vspace{-0.4em}
\textit{Insight} \circledletter[black]{3}. \textit{Scatter/gather and arithmetic should be fused and streamlined across sparse updates.}
\vspace{-0.4em}
\end{insightbox}


\subsection{Spatz Cluster}

\begin{figure}[ht]
  \vspace{-1.2em}
  \centering
  \includegraphics[width=\linewidth]{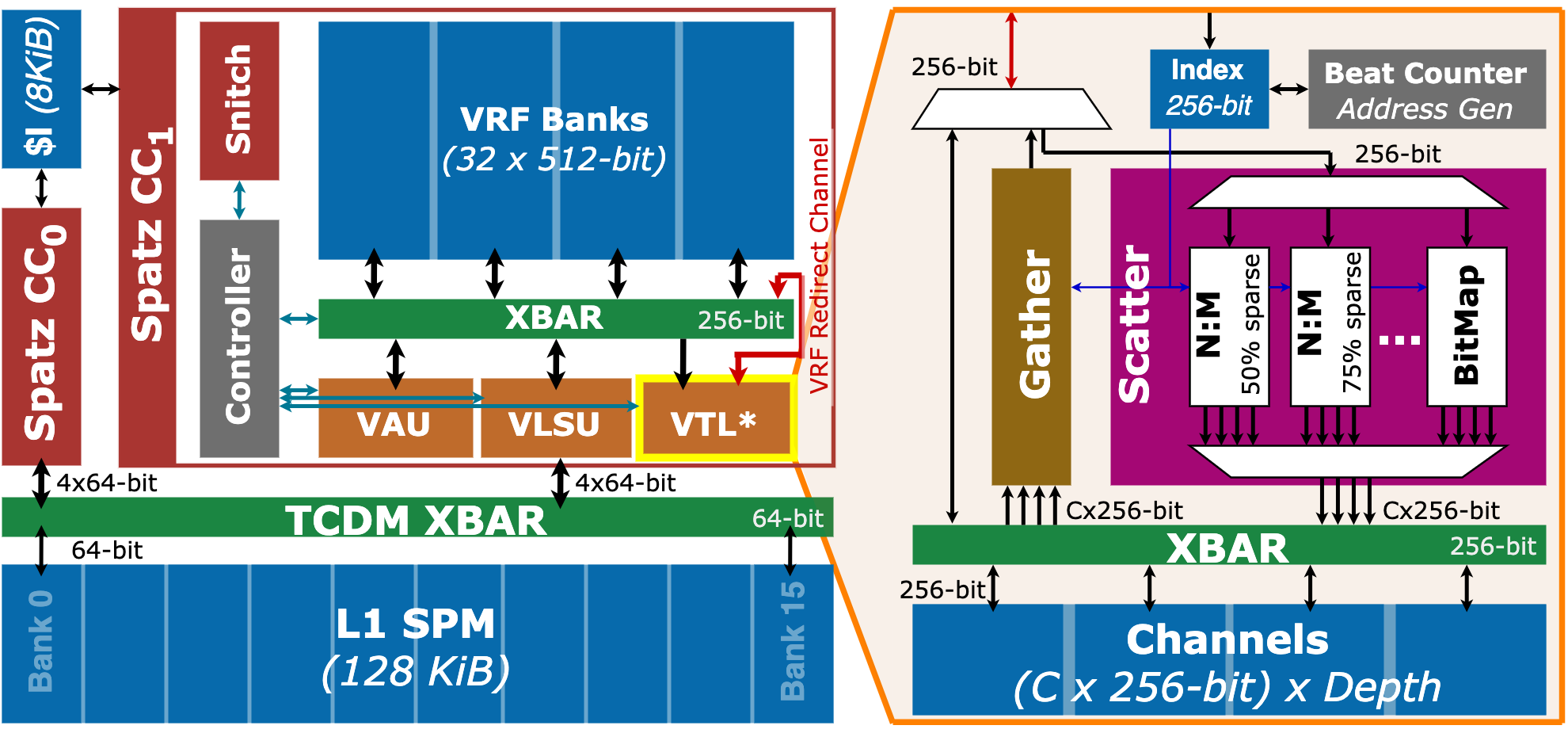}
  \vspace{-2em}
  \caption{Block diagram of the Spatz cluster with Ventaglio integration. The Ventaglio (VTL) sub-block is highlighted within the \gls{VPE}, and the inset illustrates the internal microarchitecture.}
  \label{fig:spatz_ventaglio}
    \vspace{-1.5em}
\end{figure}

Spatz~\cite{perotti2025spatz} is an open-source RVV processor designed for efficient vector computation. It is tightly coupled to a tiny single-stage RV32I Snitch core, forming a Spatz \gls{CC}. The \gls{VPE} comprises a controller, a latch-based \gls{VRF}, and parallel functional units: the \gls{VAU} and the \gls{VLSU}. The \gls{VRF} contains 32 vector registers distributed across four banks, each providing three read ports and one write port (3R1W). With a \gls{VLEN} of 512 bits, each vector register spans two consecutive banks. The \gls{VAU} executes arithmetic instructions using four parallel 64-bit \glspl{FPU}.

Two Spatz \glspl{CC} form a cluster (Fig.~\ref{fig:spatz_ventaglio}a), sharing a 128-KiB L1 \gls{SPM} through a logarithmic crossbar with single-cycle latency. In total, the cluster sustains 512 bits of FP computational throughput per cycle.

\section{Ventaglio Architecture}
\label{sec:arch}

The key challenge in sustaining full \gls{VAU} throughput in Gustavson's dataflow is a bandwidth mismatch: the \gls{VAU} operates on compact vectors, whereas accumulator state is accessed through a wider expanded window. If this state resides in \gls{VRF}, indexed reads and writes can saturate access bandwidth, delay dependent operands or results, and disrupt vector chaining. 

Ventaglio addresses these limitations by decoupling the expanded sparse state from the \gls{VRF} and storing it in a dedicated multi-channel memory unit with integrated gather/scatter support. Operating alongside the \gls{VAU}, it provides the required asymmetric bandwidth without contending for \gls{VRF} ports.

\subsection{Memory Unit Organization and Datapaths}
\label{sec:arch:mem}

The Ventaglio memory unit is implemented as a multi-channel latch-based \gls{SCM} with 1R1W ports and interleaved addressing. It comprises $C$ configurable channels, each contributing one word per cycle, such that the aggregate bandwidth scales linearly to $C\times$. Each channel is accessed independently, allowing a contiguous expanded access window to be assembled through parallel accesses across interleaved channels.

The capacity of the Ventaglio memory unit reflects a trade-off between local buffering and hardware cost. A larger unit improves latency hiding and supports deeper loop unrolling, but increases area overhead. This trade-off is particularly tight in Gustavson-style dataflow, where unrolling along the $M$ dimension requires multiple accumulation vectors to remain resident in expanded form, reducing the effective LMUL and the available unrolling budget. As a result, loop-control and address-generation overheads become harder to amortize, motivating the ISA support described in Sec.~\ref{sec:arch:integration}.

The gather datapath converts wide, multi-channel reads from the memory unit into the compact operand representation required by the \gls{VAU}, while the scatter datapath performs the inverse mapping for write-back. Ventaglio implements gather and scatter as modular datapaths, with only one sparsity mode active at a time; inactive datapaths are gated to reduce switching activity and energy. A beat counter manages packed metadata consumption and initiates preload of the next metadata word on the current word’s final valid beat.

\subsection{Spatz Integration and ISA Extensions}
\label{sec:arch:integration}

Ventaglio is integrated into Spatz as a runtime-configurable sparse execution unit operating alongside the \gls{VAU}. The active gather/scatter datapath is selected through \glspl{CSR} in the Spatz controller.

Ventaglio reuses the standard RVV vector-register namespace by dynamically mapping selected vector registers to its memory unit. A 32-bit CSR identifies the mapped registers, and a \gls{VRF} bypass channel redirects their accesses to Ventaglio instead of the standard \gls{VRF}. This preserves architectural visibility, avoids invasive scoreboard changes, and keeps indexed sparse operations transparent to the \gls{VAU}.

\begin{table}[!t]
\vspace{-1em}
\caption{Operand mapping of the proposed Ventaglio extensions.}
\vspace{-1em}
\label{tab:ventaglio-operands}
\centering
\footnotesize
\setlength{\tabcolsep}{2.5pt}
\renewcommand{\arraystretch}{1.08}

\begin{tabularx}{\columnwidth}{@{}lccccc@{}}
\toprule
\textbf{Instruction} &
\makecell{\textbf{Funct6}\\\textbf{[31:26]}} &
\makecell{\textbf{Src2}\\\textbf{[24:20]}} &
\makecell{\textbf{Src1}\\\textbf{[19:15]}} &
\makecell{\textbf{Sub-op}\\\textbf{[14:12]}} &
\makecell{\textbf{Dst}\\\textbf{[11:7]}} \\
\midrule
\texttt{vfxmacc.vf}            & \texttt{\{-, idx[4:3]\}} & wt  & act  & \texttt{idx[2:0]} & acc \\
\texttt{vlx.v}                 & --                       & --  & addr & --                & vdst \\
\midrule
\texttt{p.vle\{EEW\}.v} & --                       & inc & addr & --                & vdst \\
\texttt{p.vlx.v}               & --                       & inc & addr & --                & vdst \\
\texttt{p.fl\{b,h,w,d\}}       & --                       & inc & addr & --                & rdst \\
\bottomrule
\end{tabularx}
\vspace{-2em}
\end{table}

Table~\ref{tab:ventaglio-operands} summarizes the proposed ISA extensions. The \texttt{vlx.v} instruction loads packed index metadata from the L1 \gls{SPM} into a \gls{VRF}, while \texttt{vfxmacc.vf} performs fused indexed multiply-accumulate and selects the corresponding index context through a 5-bit \texttt{idx} field split across \texttt{funct6} and \texttt{sub-op}. Fig.~\ref{fig:kernels}(c) shows the resulting kernel implementation.

Ventaglio further extends memory operations with address-post-increment semantics \cite{schuiki2020stream}, removing explicit pointer-arithmetic instructions \circledletter[ethbronze]{D} from the critical path. The resulting streamlined kernel is shown in Fig.~\ref{fig:kernels}(d), and its performance is evaluated in Sec.~\ref{subsec:roofline}.

\section{Evaluation Methodology and Results}


We synthesized, placed and routed the Ventaglio-integrated Spatz cluster 
in a \SI{12}{\nano \meter} FinFET technology.
The reported implementation results refer to the configuration with $C=4$, which provides a total Ventaglio capacity of \SI{1}{KiB} and supports 1:4, 2:4, and bitmap sparse formats for moderate-sparsity Transformer inference.
Targeting \SI{1}{\giga \hertz} under typical conditions (TT, \SI{0.80}{\volt}, 25$^{\circ}$C), the cluster occupies \SI{4.21}{\mega GE}, corresponding to a logic-area overhead of \SI{3.1}{\percent} compared with the baseline Spatz cluster, while introducing no new critical path. Within the \gls{CC}, Ventaglio contributes \SI{9.1}{\percent} of the total area, of which \SI{83.6}{\kilo GE} (\SI{74.3}{\percent}) is attributed to the memory unit, \SI{23.4}{\kilo GE} (\SI{20.8}{\percent}) to the scatter/gather datapaths, and \SI{5.7}{\kilo GE} (\SI{4.9}{\percent}) to control logic.

\subsection{Kernel Benchmark Setup and Roofline Analysis}
\label{subsec:roofline}

We evaluate three sets of sparse tensor-contraction kernels for both
matrix--matrix and matrix--vector workloads:
\circled{1}~dense activations with \SI{50}{\percent} sparse weights in 2:4 format;
\circled{2}~dense activations with \SI{75}{\percent} sparse weights in 1:4 format; and
\circled{3}~\SI{50}{\percent} sparse activations with \SI{60}{\percent} sparse weights in bitmap format.
We compare three hardware configurations: the baseline Spatz cluster; the
Ventaglio-integrated Spatz cluster with the proposed \texttt{vfxmacc}
extension; and the same design further augmented with address-post-increment support. The representative logical kernel size is
$256(1)\times256\times256$. All kernel benchmarks are simulated using
QuestaSim~2022.3.

\begin{figure}[ht]
\vspace{-1em}
  \centering
  \includegraphics[width=\linewidth]{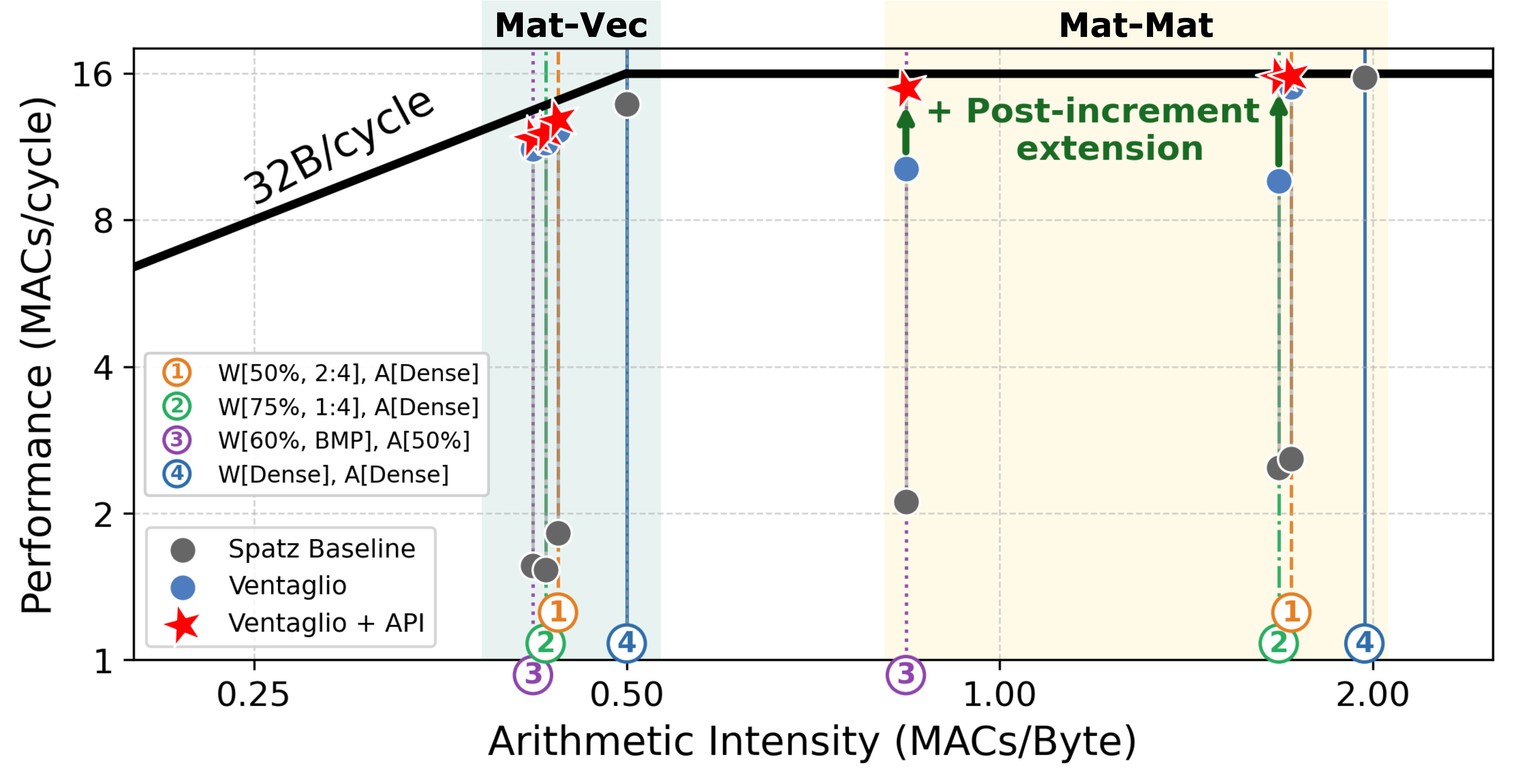}
  \vspace{-2em}
  \caption{Roofline analysis of sparse tensor contraction kernels. Circled numbers denote workload configurations.}
  \label{fig:ventaglio_roofline}
    \vspace{-1em}
\end{figure}

Presented in Fig.~\ref{fig:ventaglio_roofline}, the baseline Spatz reaches near-roofline VPU utilization for dense workloads, but performs well below the roofline on sparse
workloads due to software metadata translation and L1-backed indexed gather/scatter.
Ventaglio removes these bottlenecks by supporting metadata-driven indexed accumulation in hardware, bringing most sparse kernels close to the roofline. 

The remaining gap appears in two cases. First, as the weight matrix becomes sparser, Gustavson-style execution must maintain more output accumulation states, increasing memory-unit pressure and forcing a smaller LMUL configuration. The resulting shorter vector operations make explicit pointer updates more costly. Second, sparse activations require additional control instructions to select nonzero activation elements. Because the memory unit is capacity-constrained, these overheads cannot be fully amortized by vector operations. The address-post-increment extension reduces this residual overhead, bringing all evaluated workloads to the roofline and improving performance by $6.9$--$7.4\times$ over baseline Spatz.

\subsection{End-to-End Evaluation Setup and Results}
\begin{figure}[ht]
\vspace{-1em}
  \centering
  \includegraphics[width=\linewidth]{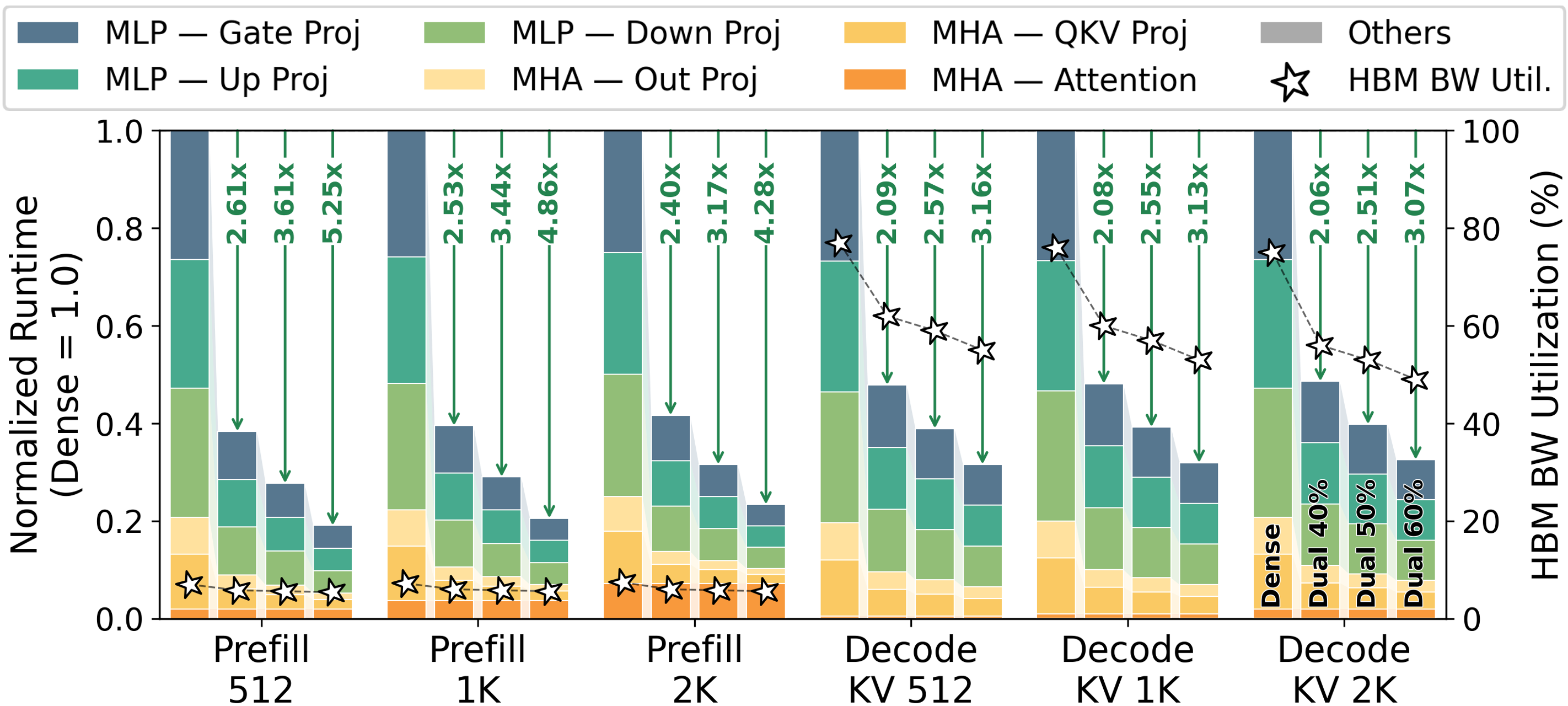}
  \vspace{-2em}
  \caption{End-to-end runtime breakdown for LLaMA-3-8B inference across prefill and autoregressive decode stages. Dense baseline is normalized to 1.0 per case. Stars indicate HBM BW utilization (right axis).}
  \label{fig:ventaglio_e2e}
    \vspace{-2em}
\end{figure}

\vspace{-0.5em}
We build a performance-accurate instruction-level model of the Ventaglio extension in GVSoC\cite{bruschi2021gvsoc}, calibrate it against the RTL implementation with a maximum performance discrepancy of 2.3\%, and scale the Ventaglio-integrated Spatz cluster to a $4\times4$ multi-cluster system connected through a 2D NoC. Each cluster interface exposes a 1024-bit network link and supports collective communication primitives, including row-wise and column-wise multicast and sum reduction. Off-chip memory is modeled in DRAMSys\cite{jung2015dramsys} as $4\times2$ HBM2 channels distributed along the west and south edges of the array, for an aggregate peak bandwidth of 512\,GB/s. We perform end-to-end inference evaluation on the LLaMA-3-8B model in FP16 precision for both prefill (S=512, 1K, 2K) and autoregressive decoding (S=1; KV=512, 1K, 2K). Fig. \ref{fig:ventaglio_e2e} compares dense execution against a DuoGPT-pruned LLaMA-3-8B model with practical 40--60\% dual-sparsity, i.e., zero-element ratios in both weights and activations.

During prefill, matrix--matrix projections dominate execution, accounting for over
95\% of the layer runtime and making the stage compute-bound. Activation and weight sparsity therefore reduce the FLOP cost directly, while Ventaglio sustains at-the-roofline efficiency on the resulting sparse kernels. This yields a $2.40$--$5.25\times$ speedup over the dense baseline.

During autoregressive decoding, matrix--vector projections are memory-bound because each cluster must load its corresponding weight tile, leaving no weight reuse across
clusters. Weight sparsity reduces HBM traffic, while the activation-sparsity-aware SUMMA dataflow transfers only the weight vectors corresponding to nonzero activation indices. However, the resulting sparse access pattern creates short, disaggregated DMA bursts that reduce bandwidth
utilization. Together, we achieve a $2.06$--$3.16\times$ speedup, translating Ventaglio-enabled at-the-roofline sparse tensor contraction and reduced HBM traffic into a sizable end-to-end acceleration.

\section{Conclusion}
\vspace{-0.5em}
Ventaglio enables efficient moderate-sparsity tensor contractions by elevating metadata-driven indexed accumulation to a native RVV execution primitive. 
Integrated into Spatz, it delivers 6.9--7.4$\times$ kernel speedup at low area cost, and achieves 2.06--5.25$\times$ end-to-end acceleration for sparse LLaMA-3 inference. 
Ventaglio shows that low-overhead hardware enhancements coupled with ISA support can deliver a significant performance and efficiency boost for sparse tensor computation.

\bibliographystyle{IEEEtran} 
\bibliography{bibliography.bib}

@IEEEtranBSTCTL{IEEEexample:BSTcontrol,
  CTLuse_forced_etal       = "yes",
  CTLmax_names_forced_etal = "3",
  CTLnames_show_etal       = "3"
}

@inproceedings{scaling_law,
  title={Scaling laws for generative mixed-modal language models},
  author={Aghajanyan, Armen and Yu, Lili and Conneau, Alexis and Hsu, Wei-Ning and Hambardzumyan, Karen and Zhang, Susan and Roller, Stephen and Goyal, Naman and Levy, Omer and Zettlemoyer, Luke},
  booktitle={International Conference on Machine Learning},
  pages={265--279},
  year={2023},
  organization={PMLR}
}

@inproceedings{yang2025wanda++,
  title={Wanda++: Pruning large language models via regional gradients},
  author={Yang, Yifan and Zhen, Kai and Ganesh, Bhavana and Galstyan, Aram and Huybrechts, Goeric and M{\"u}ller, Markus and K{\"u}bler, Jonas M and Swaminathan, Rupak Vignesh and Mouchtaris, Athanasios and Bodapati, Sravan Babu and others},
  booktitle={Findings of the Association for Computational Linguistics: ACL 2025},
  pages={4321--4333},
  year={2025}
}

@inproceedings{liu2025toward,
  title={Toward adaptive large language models structured pruning via hybrid-grained weight importance assessment},
  author={Liu, Jun and Kong, Zhenglun and Zhao, Pu and Yang, Changdi and Shen, Xuan and Tang, Hao and Yuan, Geng and Niu, Wei and Zhang, Wenbin and Lin, Xue and others},
  booktitle={Proceedings of the AAAI Conference on Artificial Intelligence},
  volume={39},
  number={18},
  pages={18879--18887},
  year={2025}
}

@article{bai2024sparsellm,
  title={{SparseLLM: Towards global pruning of pre-trained language models}},
  author={Bai, Guangji and Li, Yijiang and Ling, Chen and Kim, Kibaek and Zhao, Liang},
  journal={Advances in Neural Information Processing Systems},
  volume={37},
  pages={46203--46225},
  year={2024}
}

@inproceedings{liu2024training,
  title={Training-free activation sparsity in large language models},
  author={Liu, James and Ponnusamy, Pragaash and Cai, Tianle and Kim, Yoon and Athiwaratkun, Ben and others},
  booktitle={International Conference on Learning Representations},
  volume={2025},
  pages={98302--98322},
  year={2025}
}

@inproceedings{akhauri2024shadowllm,
  title={{ShadowLLM: Predictor-based contextual sparsity for large language models}},
  author={Akhauri, Yash and AbouElhamayed, Ahmed F and Dotzel, Jordan and Zhang, Zhiru and Rush, Alexander M and Huda, Safeen and Abdelfattah, Mohamed S},
  booktitle={Proceedings of the 2024 Conference on Empirical Methods in Natural Language Processing},
  pages={19154--19167},
  year={2024}
}

@inproceedings{ma2024first,
  title={Dynamic sparse no training: Training-free fine-tuning for sparse llms},
  author={Zhang, Yuxin and Zhao, Lirui and Lin, Mingbao and Yunyun, Sun and Yao, Yiwu and Han, Xingjia and Tanner, Jared and Liu, Shiwei and Ji, Rongrong},
  booktitle={International Conference on Learning Representations},
  volume={2024},
  pages={249--264},
  year={2024}
}

@techreport{nvidia2020ampere,
  title       = {{NVIDIA A100 Tensor Core GPU Architecture}},
  author      = {{NVIDIA}},
  institution = {NVIDIA},
  year        = {2020},
  url         = {https://images.nvidia.com/aem-dam/en-zz/Solutions/data-center/nvidia-ampere-architecture-whitepaper.pdf}
}

@online{nvidia2022hopper,
  title   = {{NVIDIA Hopper Architecture In-Depth}},
  author  = {{NVIDIA}},
  year    = {2022},
  url     = {https://developer.nvidia.com/blog/nvidia-hopper-architecture-in-depth/}
}

@online{nvidia2025blackwelltc,
  title   = {{Blackwell SM100 GEMMs}},
  author  = {{NVIDIA}},
  year    = {2025},
  url     = {{https://docs.nvidia.com/cutlass/latest/media/docs/cpp/blackwell-functionality.html}}
}

@techreport{amd2025mi300isa,
  title       = {{AMD Instinct MI300 Instruction Set Architecture Reference Guide}},
  author      = {{AMD}},
  institution = {Advanced Micro Devices},
  year        = {2025},
  url         = {https://www.amd.com/content/dam/amd/en/documents/instinct-tech-docs/instruction-set-architectures/amd-instinct-mi300-cdna3-instruction-set-architecture.pdf}
}

@online{arm2024ethosu_arch,
  title   = {{Arm Ethos-U NPU Hardware Architecture}},
  author  = {{Arm}},
  year    = {2024},
  url     = {https://developer.arm.com/documentation/109267/0103/Arm-Ethos-U-NPU/Ethos-U-hardware-architecture}
}

@article{perotti2025spatz,
  title={{Spatz: Clustering compact RISC-V-based vector units to maximize computing efficiency}},
  author={Perotti, Matteo and Riedel, Samuel and Cavalcante, Matheus and Benini, Luca},
  journal={IEEE Transactions on Computer-Aided Design of Integrated Circuits and Systems},
  volume={44},
  number={7},
  pages={2488--2502},
  year={2025},
  publisher={IEEE}
}

@inproceedings{tengfei2026scg,
  title={{RISC-V ISA Extensions for Vectorized Unstructured Sparse SpMM in LLM Inference}},
  author={Tengfei, Xia and Zhihua, Fan and Jing, Xue and Shantian, Qin and Xiaochun, Ye and Wenming, Li},
  booktitle={2026 Design, Automation \& Test in Europe Conference \& Exhibition (DATE)},
  pages={1--6},
  year={2026},
  organization={IEEE}
}

@article{titopoulos2025optimizing,
  title={{Optimizing structured-sparse matrix multiplication in RISC-V vector processors}},
  author={Titopoulos, Vasileios and Alexandridis, Kosmas and Peltekis, Christodoulos and Nicopoulos, Chrysostomos and Dimitrakopoulos, Giorgos},
  journal={IEEE Transactions on Computers},
  volume={74},
  number={4},
  pages={1446--1460},
  year={2025},
  publisher={IEEE}
}

@inproceedings{bruschi2021gvsoc,
  title={{GVSoC: a highly configurable, fast and accurate full-platform simulator for RISC-V based IoT processors}},
  author={Bruschi, Nazareno and Haugou, Germain and Tagliavini, Giuseppe and Conti, Francesco and Benini, Luca and Rossi, Davide},
  booktitle={2021 IEEE 39th International Conference on Computer Design (ICCD)},
  pages={409--416},
  year={2021},
  organization={IEEE}
}

@online{arm2025sme2_sparse_op,
  author       = {{Arm Ltd.}},
  title        = {C3.11.19 Structured sparsity outer product},
  year         = {2025},
  url          = {https://developer.arm.com/documentation/ddi0487/maa/-Part-C-The-AArch64-Instruction-Set/-Chapter-C3-A64-Instruction-Set-Overview/-C3-11-Data-processing---SME--SME2/-C3-11-19-Structured-sparsity-outer-product},
  note         = {Arm Architecture Reference Manual for A-profile architecture, accessed 2026-04-10}
}

@article{grattafiori2024llama,
  title={{The LLaMA 3 herd of models}},
  author={Grattafiori, Aaron and Dubey, Abhimanyu and Jauhri, Abhinav and Pandey, Abhinav and Kadian, Abhishek and Al-Dahle, Ahmad and Letman, Aiesha and Mathur, Akhil and Schelten, Alan and Vaughan, Alex and others},
  journal={arXiv preprint arXiv:2407.21783},
  year={2024}
}

@article{yin2025duogpt,
  title={{DuoGPT: Training-free Dual Sparsity through Activation-aware Pruning in LLMs}},
  author={Yin, Ruokai and Li, Yuhang and Lee, Donghyun and Panda, Priyadarshini},
  journal={Advances in Neural Information Processing Systems},
  volume={38},
  pages={146883--146912},
  year={2026}
}

@article{schuiki2020stream,
  title={{Stream semantic registers: A lightweight RISC-V ISA extension achieving full compute utilization in single-issue cores}},
  author={Schuiki, Fabian and Zaruba, Florian and Hoefler, Torsten and Benini, Luca},
  journal={IEEE Transactions on Computers},
  volume={70},
  number={2},
  pages={212--227},
  year={2020},
  publisher={IEEE}
}

@article{jung2015dramsys,
  title={{DRAMSys: A flexible DRAM subsystem design space exploration framework}},
  author={Jung, Matthias and Weis, Christian and Wehn, Norbert},
  journal={IPSJ Transactions on System and LSI Design Methodology},
  volume={8},
  pages={63--74},
  year={2015},
  publisher={Information Processing Society of Japan}
}

@inproceedings{dabbelt2016vector,
  title={Vector processors for energy-efficient embedded systems},
  author={Dabbelt, Daniel and Schmidt, Colin and Love, Eric and Mao, Howard and Karandikar, Sagar and Asanovic, Krste},
  booktitle={Proceedings of the Third ACM International Workshop on Many-Core Embedded Systems},
  pages={10--16},
  year={2016}
}

\end{document}